# State-selective single-electron capture in 30 keV N³⁺-He collisions

J. W. Xu[a,b], X. L. Zhu[a], W. T. Feng[a], D. M. Zhao[a], D. L. Guo[a], Y. Gao[a], S. F. Zhang[a], X Ma[a]

[a] Institute of Modern Physics, Chinese Academy of Sciences, Lanzhou 730000, China

[b] University of Chinese Academy of Sciences, Beijing 100049, China

(November 19, 2018)

The Cold-target recoil-ion momentum spectroscopy (COLTRIMS) has been employed to study the single-electron capture processes in collisions of N³⁺ ions with He atoms at an impact energy of 30 keV. The relative differential cross sections for the capture to different orbitals of N³⁺ ions are obtained and are compared with other experiments at low energy. The predictions of the molecular Columbic barrier model have been made. From the longitudinal momentum spectrum of recoil ions, the different electronic configuration was identified and the metastable projectile ions were distinguished. The single electron capture into the N²⁺ ($1s^2 2s 2p$ $^2P$) state from the ground state N³⁺ ($1s^2 2s^2$ $^1S$) projectile is the dominant reaction channel. The fraction of the metastable state N³⁺($2s2p$ $^3P$) of the incident projectile beam is about 46%.

The single-electron capture (SC) is a process in which one electron transfers from the target atom into the projectile ion. SC process has been heavily studied over past decades. It is important not only for fundamental scattering theory but also has a practical application in astrophysics, plasma and studies of biology radiation effects induced by charged ions, etc. The astronomical research indicates that solar wind ion species are dominant by proton and helium and only a fraction of 0.1% of heavy ions (such as C, N, O, Ne, Si, Fe, S and so on.) [1]. Although the abundance of heavy ions is very small, they represent the main source of x-ray, produced due to the interaction between neutral atmosphere of the planet and heavy ions. In cool and low-density interstellar medium, hydrogen and helium represent roughly 75% and 24% of the environment, respectively [2]. Cold atomic hydrogen target is not easy to handle, helium is a good candidate for our experiment.

Nitrogen ions are an important component of the solar wind, yet still many of earlier works studied electron capture cross sections with atomic and molecular hydrogen target. For N³⁺-He collisions system, few works were reported [3-13]. Gardner and Bayfield report the single-electron-capture cross sections with (4.6±1.7)×10⁻¹⁵ cm² in N³⁺-He collisions at an impact energy of 24 keV [3]. K. Ishii uses mini-EBIS facility to measure single and double electron-capture cross sections in N³⁺ with helium at projectile energies of 1.5 eV to 5.4 keV [4]. E. Y. Kamber and co-worker investigated the transfer excitation processes of single electron capture by N³⁺ ions from He using the translational energy-gain spectroscopy at energy range from 6 to15 keV. The final state populations were studied and the projectile energy dependence of the cross sections [5]. D. Burns and co-workers, in the study of state-selective electron capture by state-prepared N²⁺ ($^2P$) ions in collisions with molecular hydrogen, developed the new experimental technique of double translational energy spectroscopy to produce the pure beams of ground state N²⁺ ($2s^2 2p$ $^2P$). The single electron capture processes in collisions of 3 keV N³⁺ and helium were adopted. The measured energy gain spectrum shows that single electron capture to the $N^{2+}$ ($2s2p^2$ $^2P$) state from the metastable states $N^{3+}$ ($2s2p$ $^3P$) projectile is dominant channel [6]. In 2016, Yoh Itoh measured the relative state-selective differential cross sections for one-electron capture reactions in ground state $N^{3+}$-He collisions at $E_{lab} = 33$ eV using a crossed-beam apparatus. Only the transfer excitation processes ($N^{3+}$ ($2s^2$ $^1S$) $\rightarrow$ $N^{2+}$ ($2s2p^2$ $^2D$)), in which a single electron is captured into 2p state with simultaneous excitation of 2s electron to 2p state of the projectile were observed [7]. From the theoretical side, X. J. Liu and co-workers reported the single- and double-electron capture processes in low-energy collisions of ground state $N^{3+}$ ($2s^2$ $^1S$) with helium, where they used quantum-mechanical molecular-orbital close-coupling method [12]. The collision energy is from 0.1 eV to 15 keV. They predicted that electrons are captured to exoergic channels $N^{2+}$ ($2s2p^2$ $^2D$, $^2S$) which is the dominant mechanisms for the SEC processes. To the best of our knowledge, no work for the electron capture process in $N^{3+}$-He collisions have been reported at collision energies higher than 24 keV.

In this paper, we investigated state-selective single-electron capture processes:

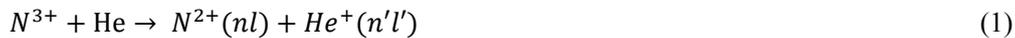
$$N^{3+} + \text{He} \rightarrow N^{2+}(nl) + He^+(n'l') \qquad (1)$$

by 30 keV $N^{3+}$ ions with He gas target. The relative state-selective differential cross sections are obtained and compared with the molecular Columbic barrier model (MCBM) [13].

## Experiment setup

The experiment was performed using the IMP EBIS-A facility of the Institute of Modern Physics, CAS in Lanzhou [14]. The standard COLTRIMS [15, 16] was used to record the recoil ion momenta from single electron capture in coincidence mode with scattered $N^{2+}$ ions. Our COLTRIMS experimental apparatus and method are described previously in [17]. The $N^{3+}$ ions produced and extracted from EBIS-A source are accelerated to an energy of 30 keV via high voltage platform. The collimated beam using the four-jaw slit-system were crossed with a supersonic Helium beam, and the scattered ions charge states were analyzed by using an electrostatic analyzer with a position-sensitive detector (PSD-P) equipped delay-line anode. The recoil target ions were guided to another position-sensitive detector (PSD-R) by an electric field perpendicular to the projectile beam and target beam. The information of the recoil ion hitting the PSD-R detector is recorded event by event. The 3-dimensional momenta can be deduced from the position and arrival time of recoil ion at PSD-R detector.

## Results and Discussion

For the single-electron capture process, the longitudinal (along the projectile beam direction) and transverse momentum (perpendicular to the projectile beam direction) of recoil $He^+$ ion was measured using COLTRIMS in coincidence with the projectile which captures one electron. The longitudinal momenta can bring information on the state of the projectile products and partial cross sections. However, the measurement of the transverse momenta can directly give the angular differential cross sections, which contains the information on the dynamics of the populated states. The longitudinal momentum $P_{long}$ and transverse momentum $P_{trans}$ of recoil ion are given by:

$$P_{long} = -\frac{Q}{v_p} - \frac{nv_p}{2} \tag{2}$$

$$P_{trans} = \theta \cdot P_0 \tag{3}$$

where Q is the binding energy change before and after the collision [17], $v_p$ is projectile velocity and $n$ the number of captured electrons; $P_0$ is initial momentum of projectile ion. All quantities in the above equations are given in atomic units.

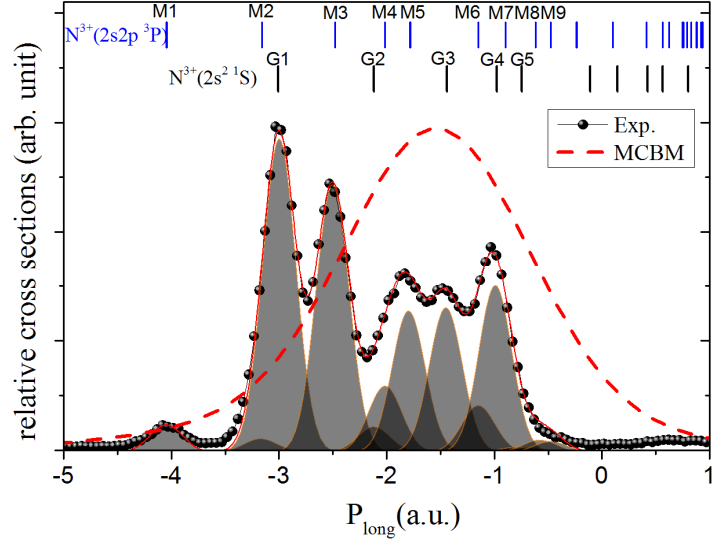

Figure 1. The longitudinal momentum of recoil He$^+$ ion for single-electron capture in 30 keV N$^{3+}$-He collisions. The red dashed line is the calculated reaction window based on the MCBM model.

Figure 1 shows the longitudinal momentum of recoil He$^+$ ion for the single-electron capture in N$^{3+}$-He collisions at an impact energy of 30 keV. The two-row vertical lines show the longitudinal momentum of single electron captured to different electronic configurations for the metastable states N$^{3+}$ (1s$^2$2s2p $^3$P) (upper lines) and ground state N$^{3+}$ (1s$^2$2s$^2$ $^1$S) (lower lines). Table 1 shows the possible reaction channels, energy defect, and corresponding longitudinal momentum value for single-electron capture by ground state N$^{3+}$ (1s$^2$2s$^2$ $^1$S) and metastable states N$^{3+}$ (1s$^2$2s2p $^3$P) ions in collisions with helium. The reaction channel indicator G and M refer to the ground state and metastable states projectile, respectively. According to the longitudinal momentum of the recoil ion, the different populated states were clearly identified. The single electron capture into the N$^{2+}$ (1s$^2$2s$^2$2p $^2$P) state (G1) from the ground state N$^{3+}$ (1s$^2$2s$^2$ $^1$S) is the dominant reaction channel, followed by single electron capture into the N$^{2+}$ (1s$^2$2s2p$^2$ $^2$P) state (M3) from the metastable state N$^{3+}$ (1s$^2$2s2p $^3$P). In addition, single electron capture into N$^{2+}$ (1s$^2$2s2p$^2$) states (G3, G4, M4, and M5) from the N$^{3+}$ projectile have a significant contribution, while the contribution of other states is rather small. The longitudinal momentum distribution is clearly different from low energy case. E. Y. Kamber [5] and co-worker studied the state-selective single electron capture at impact energies from 6 keV to 15 keV, the transfer excitation processes in which the ground state N$^{3+}$ (1s$^2$2s$^2$ $^1$S) projectile captures one electron to the $N^{2+}$ ($1s^2 2s2p^2$ $^2D$) state (G3), is the dominant channel. D. Burns et al [6] reported the energy gain/loss spectrum for single electron capture in 3 keV $N^{3+}$-He collisions. They showed that the metastable state $N^{3+}$ ($1s^2 2s2p$ $^3P$) projectile captured one electron to the $N^{2+}$ ($1s^2 2s2p^2$ $^2P$) state (M5) is predominant channel. Yoh Itoh [7] study the relative state-selective differential cross sections for one-electron capture reactions in ground state $N^{3+}$-He collisions at $E_{lab}$ = 33eV, where he observed that the single electron capture into $N^{2+}$ ($2s2p^2$ $^2D$) is predominant. Comparison of these experiments with our present results for the ground state projectile, the present experiment suggests that single electron surprisingly likely prefer to occupy the $N^{2+}$ ($2s^2 2p$ $^2P$) state.

Since the longitudinal momentum resolution is limited, we use a Gaussian function to unfold the different state's contribution. The peak maximum and full width at half maximum (FWHM) in each electronic configuration remains fixed. M1 peak represents only a single state, so the FWHM of M1 peak represents the experimental longitudinal momentum resolution with 0.36 a.u. In figure 1, the fitted results shown as the gray area and red solid line corresponds to the sum of all electronic configurations.

Table 1. Possible reaction channels for the single-electron capture by ground state $N^{3+}$ ($1s^2 2s^2$ $^1S$) and metastable states $N^{3+}$ ($1s^2 2s2p$ $^3P$) ions colliding with helium at 30 keV.

| Terms | Initial state | Final states | Q ( eV) | $P_{long}$ (a.u.) |
|-------|--------------|--------------|---------|-------------------|
| G1 | $N^{3+}(2s^2$ $^1S)$ +He($^1S$) | $N^{2+}(2s^2 2p$ $^2P)$ +He$^+(^2S)$ | 22.8 | -3.0 |
| G2 | | $N^{2+}(2s2p^2$ $^4P)$ +He$^+(^2S)$ | 15.7 | -2.1 |
| G3 | | $N^{2+}(2s2p^2$ $^2D)$ +He$^+(^2S)$ | 10.3 | -1.4 |
| G4 | | $N^{2+}(2s2p^2$ $^2S)$ +He$^+(^2S)$ | 6.6 | -1.0 |
| G5 | | $N^{2+}(2s2p^2$ $^2P)$ +He$^+(^2S)$ | 4.7 | -0.7 |
| M1 | $N^{3+}(2s2p$ $^3P)$ +He($^1S$) | $N^{2+}(2s^2 2p$ $^2P)$ +He$^+(^2S)$ | 31.2 | -4.0 |
| M2 | | $N^{2+}(2s2p^2$ $^4P)$ +He$^+(^2S)$ | 24.1 | -3.2 |
| M3 | | $N^{2+}(2s2p^2$ $^2D)$ +He$^+(^2S)$ | 18.7 | -2.5 |
| M4 | | $N^{2+}(2s2p^2$ $^2S)$ +He$^+(^2S)$ | 14.9 | -2.0 |
| M5 | | $N^{2+}(2s2p^2$ $^2P)$ +He$^+(^2S)$ | 13.1 | -1.8 |
| M6 | | $N^{2+}(2p^3$ $^4S)$ +He$^+(^2S)$ | 8.0 | -1.2 |
| M7 | | $N^{2+}(2p^3$ $^2D)$ +He$^+(^2S)$ | 6.0 | -0.9 |
| M8 | | $N^{2+}(2p^2(^1S)3s$ $^2S)$ +He$^+(^2S)$ | 3.7 | -0.61 |
| M9 | | $N^{2+}(2p^3$ $^2P)$ +He$^+(^2S)$ | 2.6 | -0.47 |

We performed the calculation of reaction window based on the MCBM model [13]. The reaction window reflects the final populated states by the single electron capture shown as a red dashed line in Figure 1. Since the MCBM model treats the projectile in its ground state. Here, we only compared with the experimental result from the ground state projectile. Figure 2 shows the comparison between the relative cross sections of the experimental results and these predicted by MCBM model. The model predicts that the G2, G3, G4, and G5 reaction channels are main reaction channels, but the experimental result indicates the G1 reaction channel is predominant near 50%. Thus G3 and G4 reaction channels have a significant contribution. This discrepancy maybe due to the fact the MCBM model is a classical model and it does not describe the quantum character of the process. In order to understand this discrepancy, a new theoretical work is needed.

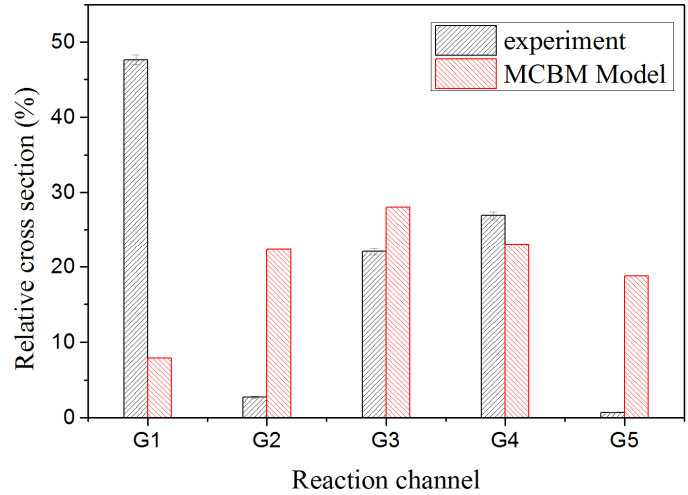

Figure 2. Comparison of relative cross sections between experimental results and MCBM model prediction, i.e. ratios of each reaction channel to the total single capture cross section versus the final populated states by captured electron.

In our experimental conditions, the ground state $N^{3+}$ ($1s^2 2s^2$ $^1S$) and metastable states $N^{3+}$ ($1s^2 2s2p$ $^3P$) in the primary projectile beam coexist. According to the experimental result, the fraction of metastable projectile is up to 46% and we assume that the single electron capture cross sections are the same for ground and metastable state $N^{3+}$ projectile. Welton et al. [11] have used a simple ionization model to compute metastable beam populations for atomic ions formed in low-density, high-electron-temperature ECR plasma-type ions sources. They computed a metastable beam fraction of 0.51 in the $N^{3+}$ ion beam, in agreement with the measurement and the estimate made by Brazuk et a l[8], while Crandall et al [10] have indicated that 10-30% of the $N^{3+}$ ions, produced in a cold-cathode Penning ion source, were in a metastable state. In our present EBIS source, the fraction of the metastable beam is consistent with the Brazuk's measurement and calculation of Welton et al.

**Conclusions**

Utilizing the COLTRIMS, we have measured the longitudinal momentum and transverse momentum of recoil $He^+$ ion for the single electron capture in collisions of $N^{3+}$ ions with He at an impact energy of 30 keV. The longitudinal momentum of recoil $He^+$ ion indicated that the dominant reaction channel is single electron capture into the $N^{2+}$ ($1s^2 2s2p$ $^2P$) state from the ground state $N^{3+}$ ($1s^2 2s^2$ $^1S$) projectile. Also, the reaction channels of transfer excitation into $N^{2+}$ ($1s^2 2s2p^2$ $^2D$, $^2S$, $^2P$) states are significant. The observed reaction channels can qualitative explained by the reaction windows from the MCBM model, but the probability of populated states have an obvious discrepancy. It's a surprise for the single electron capture from ground state at impact energies of 33 eV and 6 -15 keV, the captured electron prefers to populate $N^{2+}$ ($1s^22s2p^2\ ^2D$) state, but in the present experiments, the captured electron prefers more to occupy the $N^{2+}$ ($1s^22s^22p\ ^2P$) state. The new theoretical work is needed in order to understand this discrepancy. Also, the fraction of metastable $N^{3+}(2s2p\ ^3P)$ states in the projectile beam were up to 46%, which is consistent with the ECR sources.

**Acknowledgements**

This work is partly supported by the National Key R&D Program of China under Grant Nos. 2017YFA0402400 and 2017YFA0402300 and the Strategic Priority Research Program of the Chinese Academy of Sciences (Grant No. XDPB09-02). The authors would like to acknowledge the engineers of EBIS platform for their providing the high stability and high-quality ions beam and assistant during the experiment. XL Zhu would like to acknowledge Bennaceur Najjari for the fruitful discussion and revision of manuscript.